\documentclass[twocolumn,twocolappendix,times]{aastex63}

\usepackage{amsmath,amstext}
\usepackage[T1]{fontenc}
\usepackage{apjfonts} 
\usepackage{natbib}
\usepackage{microtype}
\usepackage{longtable}
\usepackage{verbatim}
\usepackage{CJK}

\newcommand{\hst}{{\it HST\/}}       

\newcommand{\jwst}{{\it JWST\/}}
\newcommand{\origins}{{\it Origins\/}}
\newcommand{\spitzer}{{\it Spitzer\/}}
\newcommand{\herschel}{{\it Herschel\/}}

\newcommand{\nev}{[Ne {\sc v}]~14.32~$\mu$m}
\newcommand{\oiv}{[O {\sc iv}]~25.88~$\mu$m}

\newcommand{\xray}{\hbox{X-ray}}  

\newcommand{\mstar}{M_{\star}} 
 
\newcommand{\mbh}{M_{\rm BH}} 

\newcommand{\lsix}{L_{\rm 6\mu m}}
\newcommand{\fb}{f_{\rm bulge}}
\newcommand{\lx}{L_{\rm X}}
\newcommand{\lbol}{L_{\rm bol}}
\newcommand{\fctk}{f_{\rm CTK}}

\newcommand{\fst}[1]{#1}
\newcommand{\scd}[1]{#1}

\begin{document}
\begin{CJK*}{UTF8}{gbsn}

\title[High-z BHAD]{Do current X-ray observations capture most of the black-hole accretion at high redshifts?}

\author[0000-0001-8835-7722]{Guang Yang (杨光)}
\affiliation{Department of Physics and Astronomy, Texas A\&M University, College
Station, TX, 77843-4242 USA}
\affiliation{George P.\ and Cynthia Woods Mitchell Institute for
Fundamental Physics and Astronomy, Texas A\&M University, College
Station, TX, 77843-4242 USA}

\author[0000-0001-8489-2349]{Vicente Estrada-Carpenter}
\affiliation{Department of Physics and Astronomy, Texas A\&M University, College
Station, TX, 77843-4242 USA}
\affiliation{George P.\ and Cynthia Woods Mitchell Institute for
Fundamental Physics and Astronomy, Texas A\&M University, College
Station, TX, 77843-4242 USA}
 
\author[0000-0001-7503-8482]{Casey Papovich}
\affiliation{Department of Physics and Astronomy, Texas A\&M University, College
Station, TX, 77843-4242 USA}
\affiliation{George P.\ and Cynthia Woods Mitchell Institute for
Fundamental Physics and Astronomy, Texas A\&M University, College
Station, TX, 77843-4242 USA}

\author[0000-0003-0680-9305]{Fabio Vito}
\affiliation{Scuola Normale Superiore, Piazza dei Cavalieri 7, I-56126 Pisa, Italy}

\author{Jonelle L. Walsh}
\affiliation{Department of Physics and Astronomy, Texas A\&M University, College
Station, TX, 77843-4242 USA}
\affiliation{George P.\ and Cynthia Woods Mitchell Institute for
Fundamental Physics and Astronomy, Texas A\&M University, College
Station, TX, 77843-4242 USA}

\author{Zhiyuan Yao}
\affiliation{Shanghai Astronomical Observatory, Chinese Academy of Sciences, 80 Nandan Road, Shanghai 200030, China}
\affiliation{University of Chinese Academy of Sciences, 19A Yuquan Road, Beĳing 100049, People's Republic of China}

\author{Feng Yuan}
\affiliation{Shanghai Astronomical Observatory, Chinese Academy of Sciences, 80 Nandan Road, Shanghai 200030, China}
\affiliation{University of Chinese Academy of Sciences, 19A Yuquan Road, Beĳing 100049, People's Republic of China}


\begin{abstract}
The cosmic black hole accretion density (BHAD) is critical for our understanding of the formation and evolution of supermassive black holes (BHs). 
However, at high redshifts ($z>3$), X-ray observations report BHADs significantly \fst{($\sim 10$ times)} lower than those predicted by cosmological simulations.
It is therefore paramount to constrain the high-$z$ BHAD using independent methods other than direct X-ray detections. 
The recently established relation between star formation rate and BH accretion rate among bulge-dominated galaxies provides such a chance, as it enables an estimate of the BHAD from the star-formation histories (SFHs) of lower-redshift objects. 
Using the CANDELS Lyman-$\alpha$ Emission At Reionization (CLEAR) survey, we model the SFHs for a sample of 108 bulge-dominated galaxies at $z=0.7$--1.5, and further estimate the BHAD contributed by their high-$z$ progenitors.
The predicted BHAD at $z\approx 4$--5 is consistent with the simulation-predicted values, but higher than the \xray\ measurements (\fst{by $\approx$3--10 times at $z=$4--5}).
Our result suggests that the current \xray\ surveys could be missing many heavily obscured Compton-thick active galactic nuclei (AGNs) at high redshifts.
However, this BHAD estimation assumes that the high-$z$ progenitors of our $z=0.7$--1.5 sample remain bulge-dominated where star formation is correlated with BH \fst{cold-gas} accretion. 
Alternatively, our prediction could signify a stark decline in the fraction of bulges in high-$z$ galaxies (with an associated drop in BH accretion).
\jwst\ and \origins\ will resolve the discrepancy between our predicted BHAD and the X-ray results by constraining Compton-thick AGN and bulge evolution at high redshifts.
\vspace{1.5 cm}
\end{abstract}


\section{Introduction}
\label{sec:intro}

Understanding the relation between supermassive black holes (BHs)
and their host galaxies is one of the most important tasks 
in extragalactic astronomy. 
The observations of nearby galaxies reveal a tight correlation 
with an intrinsic dispersion of $\approx 0.3$~dex
between bulge stellar mass ($\mstar$) and black hole mass 
($\mbh$; \citealt{kormendy13, saglia16}).

Tremendous observational effort has been spent to unveil 
the origin of this bulge-BH mass relation. 
One interesting clue comes from the recent observations of \cite{yang19}, which indicate that the
star formation rate (SFR) is linearly correlated with the
sample-averaged black hole accretion rate (BHAR)
among bulge-dominated galaxies at $z \approx 0.5$--2.5 (see also \citealt{kocevski17, ni19, ni21} for similar conclusions).
Their BHAR/SFR ratio ($\approx 1/300$) is similar to the observed BH-bulge mass ratio in the local universe \citep{kormendy13}.
This similarity suggests that the observed BHAR-SFR relation is strongly related to the BH-bulge connection.
Also, \cite{yang19} found that the BHAR-SFR relation does not hold for galaxies that are not bulge-dominated. 
This result indicates that BHs only coevolve with bulges rather than the disks, consistent with the observations of the local galaxies \citep{kormendy13}.
\cite{ni21} found the BHAR-SFR relation also holds for a bulge-dominated sample at lower redshift of $z \lesssim 1.2$.
Numerical simulations show that the BHAR-SFR relation is driven by fundamental accretion physics in a bulge-dominated morphological structure (Yao et al. in prep.). 
Therefore, the BHAR-SFR relation is likely a universal correlation that does not depend on redshift. 

\fst{The BHAR in \cite{yang19} was derived by averaging the X-ray detections or stacked fluxes over samples of sources (hundreds of objects per sample).
This averaging process is designed to overcome AGN short-term ($\lesssim 10^7$~years) variability and approximate long-term average BH accretion rate \citep[e.g.,][]{hickox14, yang17, yuan_f18}. 
We note that the \cite{yang19} BHAR is dominated by cold (radiative efficient) accretion rather than hot (radiative inefficient) accretion.  
This is because the average BHAR is mainly ($\gtrsim 80\%$) contributed by X-ray detected sources rather than stacking, and hot-accretion sources are below the sensitivity of the X-ray data in \cite{yang19}. 
The low BH-growth contribution from hot accretion is also expected from simulations \citep[e.g.,][]{croton06, yuan_f18}. 
Although it is observationally challenging to constrain the star-formation physical scales among the bulge-dominated galaxies, 
the simulations of bulge-dominated galaxies suggest that star formation mainly occurs on a nuclear scale of $\lesssim 1$~kpc (Yao et al. in prep.).
}

One feasible application of the BHAR-bulge SFR relation is to infer the black hole accretion density (BHAD; i.e., total BH accretion rate per comoving volume in units of $M_\odot$~yr$^{-1}$~Mpc$^{-3}$).
The BHAD, especially at high redshifts, is an important quantity for the studies of BH formation and evolution \citep[e.g.,][]{bonoli14, volonteri16}.  
However, the high-redshift BHADs from X-ray observations \citep[e.g.,][]{vito16, vito18} are significantly lower \fst{by a factor of $\sim 10$} than the theoretical results, including predictions from EAGLE \citep{crain15}, IllustrisTNG \citep{weinberger17}, and Horizon-AGN \citep{volonteri16}. 
If the simulated results are correct, then there is significantly more BH accreted mass at high redshifts and the X-ray surveys are highly incomplete, for example, because of heavy obscuration (i.e., current constraints on BH growth are missing a large fraction of highly obscured AGN at high redshifts). 
Alternatively, if the \xray\ results for BH accretion are correct, then there may be significant flaws in the simulation recipes that lead to the systematic overestimation of BH growth in the early universe. 

Therefore, it is paramount to infer the BHAD from another independent method. 
One possibility is to use constraints on the star-formation histories (SFHs) of the bulge-dominated galaxies combined with the observed BHAR-bulge SFR relation \citep{yang19}.
Recent work has shown that galaxy SFHs can be robustly constrained by modeling observed spectroscopy and photometric data with stellar population synthesis models  \citep[e.g.,][]{schreiber18, akhshik20, estrada_carpenter20}.
Here, we use \hst/WFC3 grism and broadband photometric data from the CANDELS Lyman-$\alpha$ Emission At Reionization (CLEAR) survey to constrain the SFHs of bulge-dominated galaxies for this purpose (see, \citealt{estrada_carpenter19, estrada_carpenter20}; Simons et al., in prep). 
The CLEAR fields fall in the GOODS-N and GOODS-S fields, which have deep \hst\ $H$-band imaging, 
allowing robust selection of bulge-dominated galaxies \citep{huertas_company15, huertas_company15b, yang19}.

In this work, we take the SFH results for a sample of bulge-dominated galaxies at $z=0.7$--1.5 in CLEAR.
We then estimate the BH accretion histories (BHAHs) of these individual galaxies using the observed BHAR--SFR relation from \cite{yang19}.
We sum the BHAHs and divide it by the comoving volume to estimate the redshift evolution of the cosmic black hole accretion density (BHAD) contributed by the progenitors of our bulge-dominated galaxies. 
As this BHAD history is based on only the bulge-dominated galaxies at $0.7 < z < 1.5$, it represents 
a lower bound on the total BHAD at higher redshift, as it ignores any BH growth in disk/irregular galaxies. 
We compare our predicted BHAD from the SFHs of bulge-dominated galaxies with the results from direct \xray\ observations and cosmological simulations. 

 
In this paper we use the existing constraints on galaxy SFHs and the BHAR-SFR relation for bulge-dominated galaxies to predict the BHAD at high redshifts and we use it to address the discrepancy between simulations (which predict higher BHAR at high redshifts) and X-ray surveys (that measure lower BHAR at high redshifts).     
The organization of this paper is as follows.
We present the CLEAR data analyses and our sample selection in \S\ref{sec:data}.
In \S\ref{sec:bhad}, we describe our procedures of BHAD estimation and compare 
our BHAD with the simulated and observed BHADs in the literature. 
In \S\ref{sec:discuss}, we discuss the possible uncertainties 
of different types of BHAD, present physical arguments for our results, and perform sanity checks on our results.
We summarize our work and discuss future prospects in
\S\ref{sec:sum}.

Throughout this paper, we assume a cosmology with
$H_0=70$~km~s$^{-1}$~Mpc$^{-1}$, $\Omega_M=0.3$,
and $\Omega_{\Lambda}=0.7$.
We adopt a Chabrier initial mass function (IMF;
\hbox{\citealt{chabrier03}}).
Quoted uncertainties are at the $1\sigma$\ (68\%)
confidence level, unless otherwise stated.

\section{Data and Sample}\label{sec:data}
The analyses in this work are based on the CLEAR
survey (a Cycle 23 \hst\ program, PI: C.~Papovich), which has 12
pointings of deep (12 orbit) WFC3 G102 slitless grism 
spectroscopy in the GOODS-South/North fields.
The CLEAR fields also have G141 grism and UV-to-8~$\mu$m photometric data.
The detailed data reduction and analyses of the CLEAR data are presented
in \cite{estrada_carpenter19}, \cite{estrada_carpenter20}, and Simons et al. (in prep.). 
In \S\ref{sec:sfh}, we briefly describe the modeling of CLEAR data which 
yields galaxy properties such as SFH, redshift, and $\mstar$. 
We then define the sample for this work in \S\ref{sec:sample}.

\subsection{CLEAR data modeling}
\label{sec:sfh}
\begin{figure*}
    \centering
	\includegraphics[width=2\columnwidth]{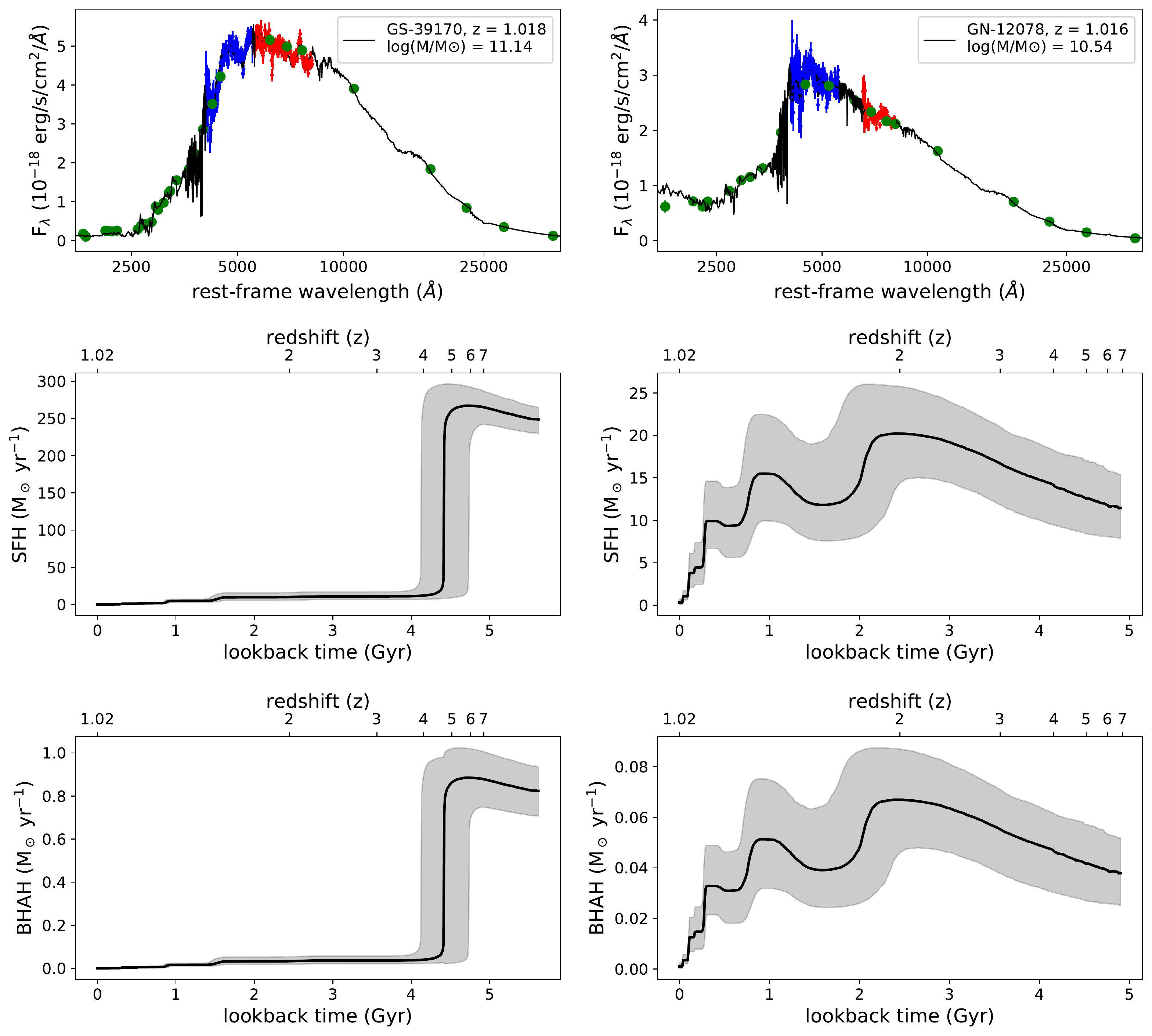}
    \caption{\fst{Example spectral fits, SFH, and BHAH for GS-39170 (left) and GN-12078 (right).
    The top panels show the best-fit spectra, the G102 grism data, the G141 grism data, and the photometry in black, blue, red, and green, respectively. 
    The middle panels show the the derived SFHs based on the spectral modelling, with uncertainties (inner 68th percentile) indicated by the shaded region.
    The bottom panels show the BHAHs from the SFHs. 
    The BHAH uncertainties (shaded region) are propagated from both of the SFH and $R$ uncertainties (see \S\ref{sec:sfh_bhad}).
    } }
    \label{fig:spec_x}
\end{figure*}

The methodology to derive the stellar populations and SFHs for our galaxies is outlined
in \cite{estrada_carpenter19, estrada_carpenter20}. For each galaxy we derive posteriors for
stellar population properties such as stellar metallicity, SFH span, SFH bins, redshift, $A(V)$, and 
$\mstar$.\footnote{The CLEAR $\mstar$ have a systematic offset of $+0.27$~dex compared to the 
CANDELS $\mstar$ \citep{santini15, barro19}.
Such a systematic is common among different codes (see, e.g., Appendix~A of \citealt{ni21}).
Since the BHAR-SFR relation in \cite{yang19} was derived based on the CANDELS SED-fitting results,  
we scale the CLEAR $\mstar$ (as well as the SFHs) down by 0.26~dex to eliminate the systematic.   
}

For our SFHs we used the flexible approach outlined in 
\cite{leja19}, wherein SFHs are modeled using a set of time bins and the mass generated in each time bin
is fit for. The maximum span of our SFHs is determined by redshift and is set to be the age of the universe.
We allow the overall span of the time bins to vary which produces smoother SFHs where as keeping the bins set would
produce a step-wise SFH. The amount of time bins used depends on the UVJ color-color diagram classification of
the galaxy, where quiescent selected galaxies use 10 time bins, and star-forming use 6. This was done because
quiescent galaxies form most of their mass early on and using more time bins would allow for a higher temporal
resolution at early formation times, while  star-forming galaxies will tend to form more of their mass later
therefore there is less of a need for higher temporal resolution at early times (this choice also reduces the
run time of our SED fits). For our sample, we use a continuity prior \citep{leja19}. This prior weighs 
towards a more continuous SFH and against a bursty history. 

The resulting SFHs were derived by sampling the 
posteriors of the SFH span, SFH bins, and stellar mass and generating 5000 iterations of SFHs. From this sample 
of we can derive our SFH (as the 50th percentile) and errors on the SFHs (inner 68th percentile).
Fig.~\ref{fig:spec_x} displays example spectra and SFH fits for two sources in our sample (\S\ref{sec:sample}).
The details of the CLEAR catalog will be presented in Simons et al.\ (in prep.). 

\subsection{Sample selection}
\label{sec:sample}
Among the CLEAR objects, we select bulge-dominated 
galaxies based on the CANDELS machine-learning $H_{160}$ morphological 
classifications \citep{huertas_company15}, following the 
same criterion as in \cite{yang19}.\footnote{These is 
one CLEAR pointing outside the CANDELS region.
In this work, we discard this pointing where morphological 
classifications are not available.}
The machine-learning-selected bulge-dominated galaxies have round 
and smooth shapes upon visual inspections (see Fig.~2 in 
\citealt{yang19}) and tend to be compact upon profile fitting
(see Fig.~C4 of \citealt{ni21}).
It is critical to select bulge-dominated galaxies, as BHAR
is only correlated with the SFR in bulge-dominated galaxies 
not the SFR in other types of galaxies \citep{yang19}.

The CLEAR catalog provides redshifts, $\mstar$, and SFHs (\S\ref{sec:sfh}).
We focus on the redshift range of $z=0.7$--1.5. 
This redshift range guarantees that the grism spectroscopy
(0.8--1.7~$\mu$m) covers important age and metallicity indicators 
of, e.g., H$\beta$, Mg$b$, and H$\alpha$. 
We select all bulge-dominated galaxies with stellar mass above
$\mstar = 10^{10}\ M_\odot$, which is the CLEAR mass limit
at $z\approx 1.5$.
Our volume-limited sample has 108 bulge-dominated objects. 

\section{Estimation of Black Hole Accretion Density}
\label{sec:bhad}

\subsection{From SFHs to BHAD}
\label{sec:sfh_bhad}

\begin{figure*}
    \centering
	\includegraphics[width=2\columnwidth]{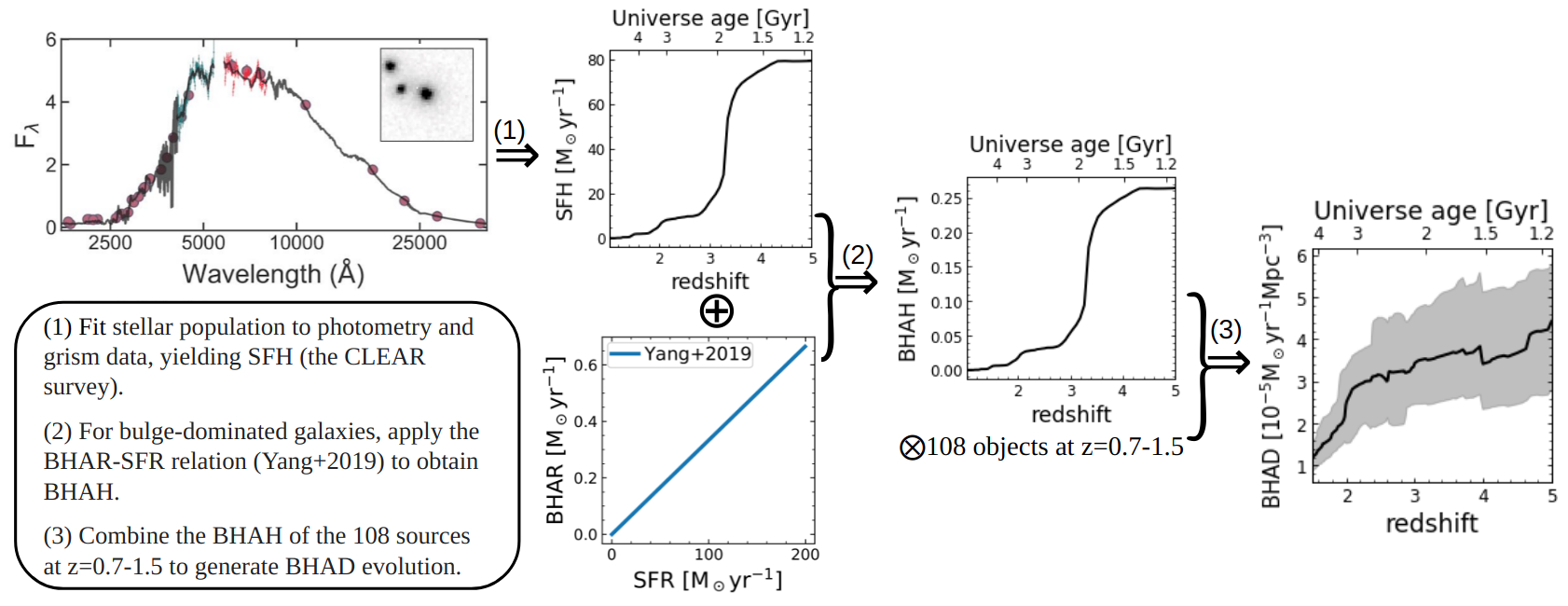}
    \caption{The schematic plot describing our procedures to estimate the BHAD. The three major steps are marked and explained at bottom left.  These are discussed in more detail in \S~\ref{sec:bhad}.
    }
    \label{fig:scheme}
\end{figure*}

\begin{figure*}
    \centering
	\includegraphics[width=2\columnwidth]{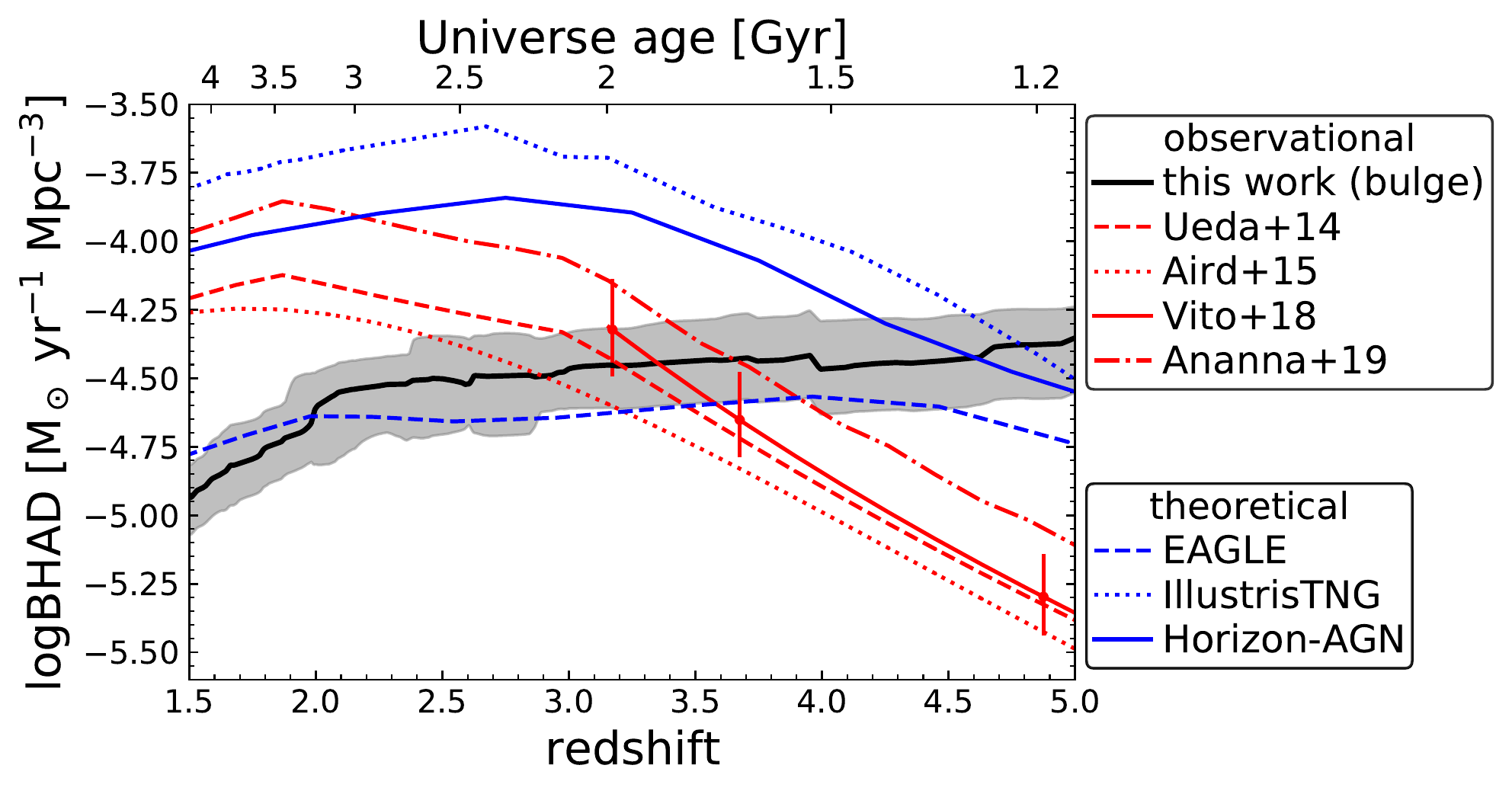}
    \caption{Black hole accretion density as a function of 
    Universe age. 
    The black curve indicates our estimated BHAD contributed
    by bulge-dominated galaxies. 
    The grey shaded region indicates 2$\sigma$ uncertainties. 
    The red and blue curves represent observational (X-ray)
    and theoretical results from the literature, respectively.
    \fst{All of the X-ray BHADs are derived assuming the same bolometric correction and radiation efficiency as adopted by \cite{yang19}.
    The red error bars represent the $1\sigma$ bootstrap uncertainties from \cite{vito18}.
    We expect similar uncertainties on the other X-ray BHADs.
    }
    At high redshifts ($z\approx 4$--5), our BHAD is similar
    to the simulation predictions, but  higher than the X-ray
    measurements. 
    }
    \label{fig:bhad}
\end{figure*}

Fig.~\ref{fig:scheme} shows the schematic for the steps we applied to convert the SFHs of individual bulge-dominated galaxies to an estimate of the BHAD.  
We take the SFH and the associated uncertainties for each galaxy derived by modeling the grism spectroscopy and broad-band photometric data (\S\ref{sec:data}). 
This is ``step 1'', labeled (1) in Fig.~\ref{fig:scheme}.  

In ``step 2'', labeled (2) in Fig.~\ref{fig:scheme}, we multiply the SFHs by the BHAR-SFR relation from \citet{yang19}. 
\fst{This procedure yields a BH accretion history (BHAH) for each object in our sample, i.e., 
\begin{equation}
\label{eq:bhah}
{\rm BHAH}_i = {\rm SFH}_i \times R,
\end{equation}
where $R=10^{-2.48}$ is the BHAR/SFR ratio derived by \citet[][see their Eq. 6]{yang19} and the subscript ($i$) represents the source index in our bulge-dominated sample.
We remind the reader that the BHAH estimated here represent the long-term average accretion rate dominated by cold accretion over the cosmic history (see \S\ref{sec:intro}).
Fig.~\ref{fig:spec_x} shows two example BHAHs and the associated uncertainties. 
The BHAH uncertainties are propagated from both of the SFH and $R$ uncertainties, using the standard error-propagation formula based on Eq.~\ref{eq:bhah}, i.e., 
\begin{equation}
    \delta {\rm BHAH}_i = {\rm BHAH}_i \sqrt{ 
    \left(\frac{\delta {\rm SFH}_i}{{\rm SFH}_i}\right)^2 +
    \left(\frac{\delta R}{R}\right)^2},
\end{equation}
where $\delta R/R= 0.05 \times \ln 10=0.12$ is the fitting uncertainty in \citep{yang19}, under the assumptions of radiation efficiency and bolometric correction. 
We address the systematic uncertainties arising from these assumptions in \S\ref{sec:xlf_bhad}.   
The SFH uncertainties are from our modeling of CLEAR data (see \S\ref{sec:sfh}).
We apply Eq.~\ref{eq:bhad_err} to upper and lower uncertainties, respectively, because the upper and lower SFH uncertainties are not always equal (e.g., Fig.~\ref{fig:spec_x}). 
}

In ``step 3'', labeled (3) in Fig.~\ref{fig:scheme}, we sum the BHAHs for the 108 bulge-dominated galaxies and divide it by the comoving volume ($V_c$) of $z=0.7$--1.5 covered by the CANDELS/CLEAR area (61~arcmin$^2$), 
\fst{i.e.,
\begin{equation}
\label{eq:bhad}
    {\rm BHAD} = \frac{\sum_{i=1}^{108}{\rm BHAH}_i}{V_c} \\
               = \frac{\sum_{i=1}^{108}{\rm SFH}_i}{V_c} R,
\end{equation}
where we apply Eq.~\ref{eq:bhah}.
We propagate the SFH and $R$ uncertainties into the BHAD using the standard error-propagation formula, i.e., 
\begin{equation}
\label{eq:bhad_err}
    \delta {\rm BHAD} = {\rm BHAD} 
    \sqrt{ 
    \frac{\sum_{i=1}^{108} (\delta {\rm SFH}_i)^2}
         {({\sum_{i=1}^{108} \rm SFH}_i)^2} +
    \left(\frac{\delta R}{R}\right)^2
    }.
\end{equation}
} 


The resulting bulge BHAD and its 2$\sigma$ error as a function of redshift/Universe age is displayed in Fig.~\ref{fig:bhad}. 
In this work, we do not extend beyond $z=5$, because there is only $\approx 1$~Gyr cosmic time at $z>5$ and our SFH measurements may not have the time resolution sufficiently high to probe the detailed SFH evolution in the first $\approx 1$~Gyr (\S\ref{sec:sfh}, see \citealt{estrada_carpenter19,estrada_carpenter20}). 
We will discuss the SFH/BHAD evolution at $z>5$ in a future dedicated work.

Fig.~\ref{fig:bhad} also displays theoretical BHADs from cosmological simulations. 
The Horizon-AGN BHAD curve \citep{volonteri16} is compiled by \cite{vito18}. 
The EAGLE \citep{crain15, schaye15} and IllustrisTNG \citep{weinberger17, pillepich18} results are from the corresponding database, where we use the simulation sets of ``RefL0100N1504'' (EAGLE) and ``TNG100-1'' (IllustrisTNG).
When deriving the simulated BHAD at a given redshift, we add up the BHARs from different galaxies and divide them by the simulated comoving volume. 

We remind the reader that our BHAD only accounts for BH growth in bulge-dominated galaxies with $\mstar > 10^{10}\ M_\odot$.
Since active galactic nuclei (AGNs) can also be found in 
less massive bulges as well as in non-bulge-dominated 
galaxies \citep[e.g.,][]{yang19}, our estimated BHAD is 
an lower bound for the total BHAD, i.e., any BHAD measurement similar or above our values should be considered as consistent with our estimation. 
Therefore, our BHAD is consistent with the simulated results at $z=1.5$--5 in general (see Fig.~\ref{fig:bhad}).


\subsection{BHADs based on literature X-ray luminosity functions}
\label{sec:xlf_bhad}
\fst{In this section, we derive BHADs based on the measured X-ray luminosity functions (XLFs) from the literature \citep[i.e.,][]{ueda14, aird15, vito18, ananna19}, and compare the results with our SFH-based BHAD. 
We do not use the BHADs from the literature directly, because those are estimated based on different assumptions of radiation efficiencies and bolometric corrections.
Below, when deriving the XLF-based BHADs, we adopt the same radiation efficiency and bolometric correction used by \cite{yang19} for the BHAR-SFR relation. 
In this way, we effectively address the systematic uncertainties due to radiation efficiency and bolometric correction.
}

\fst{To derive the BHAD from each XLF, we first convert the XLF to the bolometric luminosity function (BLF), i.e., 
\begin{equation}
\label{eq:blf}
\frac{d n}{d \log \lbol} = 
    \frac{d n}{d \log \lx} \frac{d \log \lx}{d \log \lbol},
\end{equation}
where $d n/d \log \lbol$ and $d n/d \log \lx$ are the BLF and XLF, respectively, and $d \log \lx/d \log \lbol$ is the derivative of the luminosity-dependent bolometric correction from \cite{hopkins07}, which is also adopted by \cite{yang19}.\footnote{\fst{\cite{yang19} scaled down the \cite{hopkins07} bolometric correction by a factor of 0.7 (see Footnote 5 of \cite{yang19} for explanation). Here, we also adopt this scaling factor.}} 
From the BLF, we can calculate the BHAD by
\begin{equation}
\label{eq:blf_bhad}
{\rm BHAD} = \frac{1-\epsilon}{\epsilon c^2} 
\int_{43}^{49} \lbol \frac{dn}{d\log \lbol} d\log\lbol,
\end{equation}
where $c$ is the speed of light and $\epsilon$ is the radiation efficiency, and the integral limits 43 and 49 [$\log(\rm erg\ s^{-1})$] correspond to $\log L_X= 42$ and 47 [$\log(\rm erg\ s^{-1})$] under the \cite{hopkins07} bolometric correction.
We adopt $\epsilon=0.1$, which is the value used by \cite{yang19}.
Fig.~\ref{fig:bhad} displays the resulting XLF-based BHADs.
These BHADs are consistent with our BHAD at low redshifts ($z\lesssim 3$), but are lower than our values by a factor of $\approx 3$--10 at $z\approx 4$--5.
}

\fst{In Fig.~\ref{fig:bhad}, we also show the uncertainties of \cite{vito18} BHAD.
These uncertainties were calculated by \cite{vito18} based on binned high-$z$ AGNs, employing a bootstrap technique. 
This bootstrap technique properly accounts for statistical fluctuations due to limited AGN sample sizes at different luminosities. 
It also propagates different types of errors, including photometric redshift, column density ($N_{\rm H}$), and X-ray fluxes. 
From Fig.~\ref{fig:bhad}, these uncertainties are small compared to the difference between \cite{vito18} BHAD and our BHAD at $z\approx 4$--5. 
Therefore, the XLF uncertainties cannot explain the discrepancy between our BHAD and the X-ray results. 
The uncertainties of the other XLF-based BHADs \citep{ueda14, aird15, ananna19}, although not publicly available,\footnote{\fst{We note that it is not feasible to propagate the XLF parametric uncertainties to the BHAD uncertainties.  
Because the XLF is determined by multiple model parameters, the BHAD uncertainties are affected by both of the variances of each single XLF parameter and the covariances of different parameters.
However, the covariances are not publicly available.}} 
should be comparable to the \cite{vito18} uncertainties at high redshifts.
This is because the BHAD uncertainties in all of the X-ray works are dominated by the relatively small number of high-$z$ AGNs detected in the existing deep X-ray surveys (e.g., CDF-S and CDF-N).
However, we caution that all of the XLF works could suffer from significant systematic uncertainties due to Compton-thick AGNs ($N_{\rm H}\gtrsim 10^{24}$~cm$^{-2}$), which are largely missed in X-ray surveys especially at high redshift. 
We discuss this issue in \S\ref{sec:discuss}.
}

\section{Discussion}
\label{sec:discuss}

\subsection{Possible causes of the BHAD discrepancy}
\label{sec:err}
\begin{figure}
    \centering
	\includegraphics[width=\columnwidth]{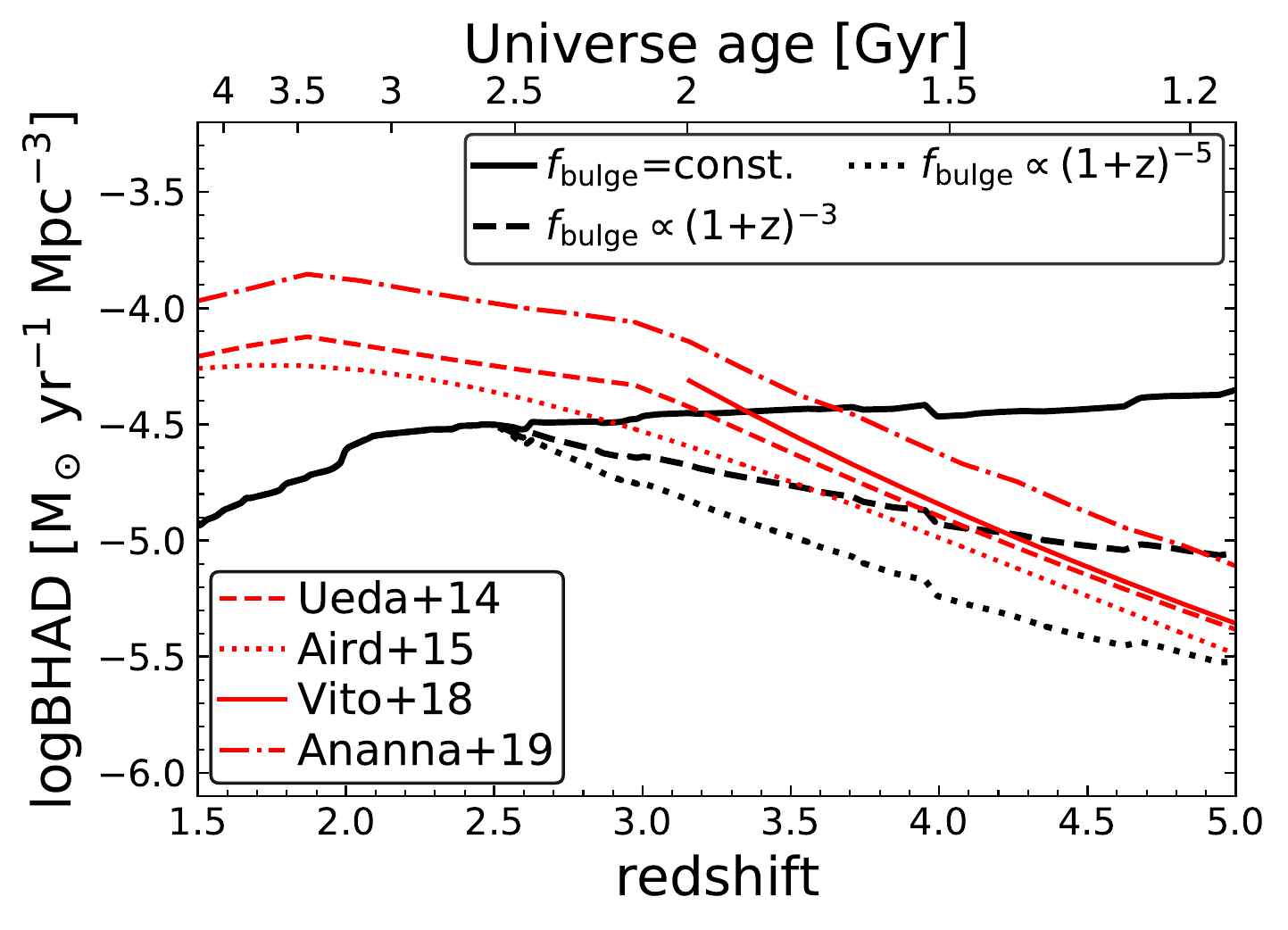}
    \caption{Same format as Fig.~\ref{fig:bhad_moduled} but 
    showing different cases of the high-redshift ($z>2.5$) 
    evolution of the bulge-dominated galaxy fraction 
    ($\fb$). 
    The solid curve assumes that $\fb$ does not evolve
    at $z>2.5$, following the trend at lower redshifts. 
    The dashed and dotted black curves assume that $\fb$ 
    decreases at $z>2.5$, following $(1+z)^{-3}$ and 
    $(1+z)^{-5}$, respectively. 
    \scd{Compared to the original BHAD (solid black curve), 
    the modulated BHADs (dashed and dotted black curves) are 
    more consistent with the results from X-ray observations 
    at $z\approx 4$--5.} 
    Therefore, if the \xray\ BHADs are physical, then $\fb$
    must evolve strongly in the early universe. 
    }
    \label{fig:bhad_moduled}
\end{figure}

\fst{Fig.~\ref{fig:bhad} shows that our BHAD at $z\approx 4$--5 is similar to theoretical predictions, but is $\approx$3--10~times higher than the \xray\ results.
This high-$z$ BHAD difference is beyond the expectations from the uncertainties of different BHADs (\S\ref{sec:bhad}).
We note that hot radiative-inefficient accretion should not be responsible for this BHAD discrepancy, because both of our BHAD and the X-ray BHADs should be dominated by cold accretion (see \S\ref{sec:intro}).
We discuss some possible causes of the discrepancy below.}

X-ray observation is often reliable in AGN selection owing to the nearly universal \xray\ emission from AGN and the weak contamination from host galaxies \citep[e.g.,][]{brandt15}.
However, AGNs could be missed by \xray\ surveys due to strong obscuration. 
\cite{vito16} performed an \xray\ stacking analysis for 
high-$z$ \xray\ undetected galaxies using the deepest \xray\ 
survey, \hbox{CDF-S} \citep{luo17}.
They found the stacked \xray\ emission is negligible compared
to that from \xray\ detected AGNs. 
Their result suggests that, if a large number of high-$z$ AGNs 
are missed, they must be heavily obscured, likely at the 
Compton-thick level (e.g., \citealt{hickox18}).

AGN obscuration generally increases toward high redshift \citep[e.g.,][]{hasinger08, liu17}. 
The fraction of obscured AGNs (both Compton-thick and Compton-thin AGN) has been modeled as a positive function of redshift, but the ratio of Compton-thick and Compton-thin AGNs ($\fctk$) is often assumed to be $\approx$ constant due to the lack of observational constraints, \fst{especially at $z\gtrsim 3$ \citep[e.g.,][]{ueda14, aird15, buchner15, ananna19}}. 
The BHAD curves of \cite{ueda14}, \cite{aird15}, and \cite{ananna19} in Figs.~\ref{fig:bhad} and \ref{fig:bhad_moduled} include the contribution from Compton-thick AGNs based on the assumption that $\fctk$ is constant. 
These BHAD estimations are still  lower at $z \approx 4$--5 than the BHAD we infer from the galaxy SFHs, suggesting that \fst{the strong assumption of a constant $\fctk$ may be incorrect, especially at high redshifts}. 
Therefore, the population of Compton-thick AGN may be dominant over the Compton-thin population at high redshifts.  
This is also supported by simulations, since the simulated BHADs are also higher than the X-ray BHADs at $z\approx 4$--5 (Fig.~\ref{fig:bhad}).
\fst{We present some physical arguments for a higher intrinsic BHAD than X-ray observed and make practical predictions for future observations in \S\ref{sec:argue}}.

In our calculation of the BHAD, we assume that the BHAR-SFR relation still holds 
for the progenitors of our bulge-dominated galaxies. 
Our BHAD prediction would be overestimated if many of the galaxy progenitors were non-bulge-dominated at earlier times \citep[e.g.,][]{kocevski17, ni21}, which could be a result of morphological transformation
in star-forming disk galaxies, caused by, e.g., major mergers. 
As one counter example, \cite{huertas_company15b} found that the bulge-dominated 
fraction ($\fb$) among the $\mstar \approx 10^{11.2}~M_\odot$ ($z=0$) 
galaxies' progenitors is roughly a constant ($\approx 20\%$--30\%) 
at $z\approx 0$--2.5, suggesting that morphological transformation 
between bulge-dominated and other types is not prevalent at least at these redshifts
(see, e.g., \citealt{mortlock13} and \citealt{conselice14}  
who arrive at similar conclusions).
However, it is still possible that such transformation happens
frequently at $z\gtrsim 2.5$. 
Probing this possibility is beyond the capability of 
current facilities due to the lack of $\gtrsim 1.6\ \mu$m high-resolution imaging, but will be testable with \jwst\ imaging. 

We quantify the effects of possible high-redshift $\fb$ evolution based on the assumption that our BHAD declines following ${\rm BHAD} \propto \fb \propto (1+z)^{-\gamma}$ at $z > 2.5$. 
In Fig.~\ref{fig:bhad_moduled} we plot the BHAD evolution curves in Fig.~\ref{fig:bhad_moduled} for different values of $\fb$.
From Fig.~\ref{fig:bhad_moduled}, to match the BHAR inferred from X-ray surveys would require a very steep power-law index of $\gamma \approx 3$--5.
This means that, if the \xray\ BHADs are correct, then $\fb$ must drop decline by a large factor of $\approx 10$--50 from $z\approx 2.5$ to $z\approx 5$.  
This provides a testable prediction for \jwst\ observations of galaxies at these redshifts. 

\subsection{Physical arguments for a higher intrinsic BHAD than X-ray observed}
\label{sec:argue}
\begin{figure}
    \centering
	\includegraphics[width=\columnwidth]{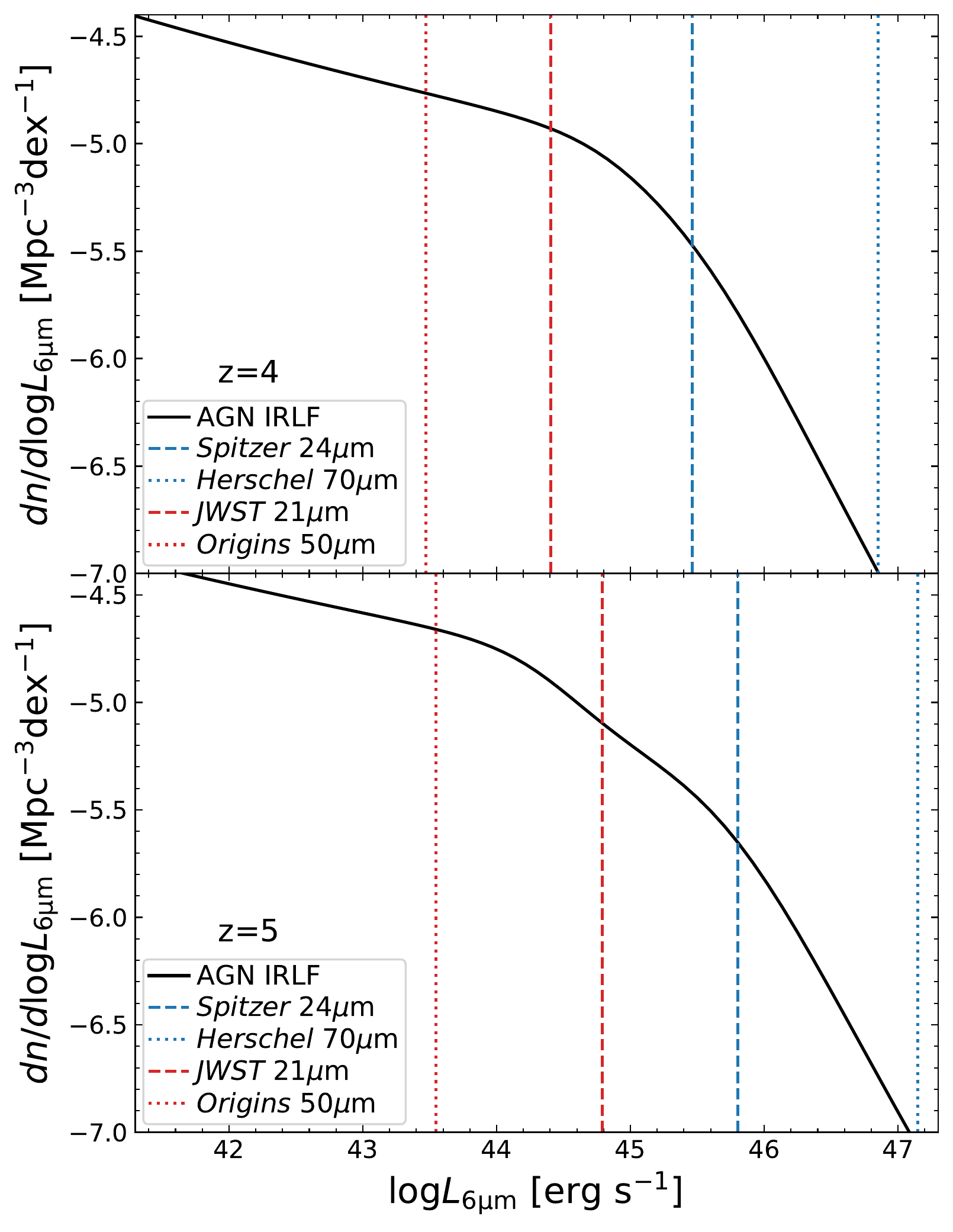}
    \caption{\fst{Predicted AGN IR luminosity ($\lsix$) function based on our BHAD at $z=4$ (top) and $z=5$ (bottom). 
    The vertical lines represent $5\sigma$ sensitivity from a typical exposure of 1000~seconds for some IR telescopes as labeled.
    \jwst\ and \origins\ will be able to sample around or below the break luminosity ($\sim 10^{44.5}$~erg~s$^{-1}$).
    }
    }
    \label{fig:irlf}
\end{figure}

Our SFH-based BHAD and the simulated BHADs are both higher than the X-ray results at $z \approx 4$--5 (Fig.~\ref{fig:bhad}). 
It is understandable that simulations predict a relatively strong BH accretion process at high redshifts, because cold gas, which fuels both star formation and AGN, is likely abundant and concentrated in the early universe. 

At high redshifts, large amounts of dust associated with the gas can totally obscure the (rest-frame) UV light from intensive star formation. 
The recent development of sub-millimeter surveys begin to reveal a large populations of heavily obscured star-forming galaxies at $z \gtrsim 3$ that are faint/undetected in shorter-wavelengths surveys \citep[e.g.,][]{gonzalez_lopez20, smail21}.   
This new population could contribute a significant (or even dominant) fraction of the cosmic star-formation rate density (SFRD; e.g., \citealt{wang19, gruppioni20}).

Likewise, the gas/dust-rich environment at high redshifts could also obscure high-$z$ AGN activity.  
If Compton-thick obscuration is common at high redshifts, then the X-ray selected AGNs will be highly incomplete, leading to an underestimation of BHAD (\S\ref{sec:err}).
From X-ray spectral analyses \citep[e.g.,][]{vito18, li19}, most ($\approx 80\%$--90\%) of the X-ray selected $z\gtrsim 3$ AGNs are Compton-thin or unobscured, and none of the detected Compton-thick AGNs have $N_{\rm H}>10^{25}$~cm$^{-2}$.
This absence of $N_{\rm H}>10^{25}$~cm$^{-2}$ AGNs is likely a selection effect due to their X-ray faintness \citep[e.g.,][]{hickox18}.
These results indicates that current X-ray selections could indeed miss many high-$z$ Compton-thick AGNs (especially at $N_{\rm H}>10^{25}$~cm$^{-2}$).
Our SFH-based and the simulated BHADs \fst{(both dominated by cold accretion; \S\ref{sec:intro})} are consistent with this interpretation: if these results are correct then it implies a large population of heavily obscured AGN at $z\gtrsim 3$ than currently found in X-ray surveys. 

\scd{Since the missed AGNs are undetected by the currently deepest X-ray surveys (i.e., CDF-S and CDF-N), their apparent (uncorrected for obscuration) luminosities must lie below the survey sensitivity ($\lx \sim 10^{42.5}$~erg~s$^{-1}$ at $z\approx 4$--5; e.g., \citealt{vito18}).}
\fst{Although these missed high-$z$ Compton-thick AGNs have weak or none X-ray signals due to heavy obscuration \cite[e.g.,][]{vito16}, but they are likely luminous at IR wavelengths due to dust re-emission.  
Therefore, they can be detected by IR telescopes. 
It is thereby useful to quantitatively predict the high-$z$ AGN IR luminosity function (IRLF) based on our SFH-based BHAD.}

\fst{To perform this task, we first take the XLF from \cite{aird15}.
We then normalize the XLF at a given redshift so that the corresponding BHAD (integrated over $\log L_X = 41$--47) equals to unity (see \S\ref{sec:xlf_bhad} for the detailed process). 
We multiply the normalized XLF by our BHAD (Fig.~\ref{fig:bhad}). 
The above procedure yields a XLF that can produce our BHAD.
We then convert this XLF to the AGN IRLF using a relation between $L_X$ and $\lsix$ (AGN 6~$\mu$m $\nu L_\nu$ luminosity; e.g., \citealt{stern15}), i.e., 
\begin{equation}
\label{eq:irlf}
\begin{split}
\frac{d\Phi}{d\log \lsix} = \frac{d\Phi}{d\log L_X} \frac{d\log L_X}{d\log \lsix} \\
    = \frac{d\Phi}{d\log L_X} \times (1.024-0.094\log(\lsix/10^{41}\ {\rm erg\ s^{-1}}) ).
\end{split}
\end{equation}
We display the resulting IRLF at $z=4$ and $z=5$ on Fig.~\ref{fig:irlf}.
We mark sensitivities of some current and future IR missions on Fig.~\ref{fig:irlf}.
These $\lsix$ limits are converted from the flux-density sensitivities for a typical 1000-second exposure, assuming a K correction based on an AGN IR spectral template generated by {\sc x-cigale} \citep{boquien19, yang20}. 
{\sc x-cigale} employs a clumpy torus model, {\sc skirtor} \citep{stalevski12, stalevski16}. 
We set the viewing angle to 70$^\circ$, typical for obscured AGNs \citep{yang20}, and leave other parameters as the default values. 
Our conclusion below is not sensitive to the IR model parameters in {\sc x-cigale}.
}

\fst{From Fig.~\ref{fig:irlf}, \spitzer\ and \herschel\ can only sample $\lsix$ more than $\approx 10$~times above the IRLF break luminosity ($L^* \sim 10^{44.5}$~erg~s$^{-1}$).
This means that \spitzer\ and \herschel\ are not able to effectively detect the predicted Compton-thick AGNs, as the IRLF declines sharply above $L^*$.
For a CANDELS-like deep survey ($\sim 1000$~arcmin$^2$), \spitzer\ (\herschel) can only detect $\approx 2$ (0) objects according to the IRLF in Fig.~\ref{fig:irlf}. 
Also, the task of AGN identification is challenging for \spitzer, as there is a large wavelength ``gap'' between the coverages of the IRAC~8$\mu$m and the MIPS~24$\mu$m filters \citep[e.g.,][]{yang21}.
The future missions of \jwst\ and \origins\ can sample $\lesssim L^*$ objects (see Fig.~\ref{fig:irlf}) thanks to their unprecedented sensitivities.
\origins\ perform better than \jwst\ because the AGN SED peak (rest-frame $\approx 5$--20~$\mu$m) is out of the \jwst\ coverage at $z\approx 4$--5.
For a CANDELS-like deep survey ($\sim 1000$~arcmin$^2$), \jwst\ (\origins) can detect $\approx 20$ ($\approx 60$) objects according to the IRLF in Fig.~\ref{fig:irlf}.
Thanks to the continuous wavelength coverage of \jwst\ and \origins, the AGN identification will be practically feasible \citep[e.g.,][]{yang21}.
}

\fst{We caution that the IRLF in Fig.~\ref{fig:irlf} assumes that Compton-thick AGNs follow the same intrinsic (obscuration-corrected) $\lx$ distribution as Compton-thin and unobscured AGNs at high redshifts. 
This assumption can be tested in the future using the Compton-thick samples detected by \jwst\ and \origins\ as above.}
\scd{If this assumption turns out to be incorrect, the Compton-thick intrinsic $\lx$ distributions can be inferred from the measured IR luminosities by \jwst\ and \origins.}

\section{Summary and Future Prospects}
\label{sec:sum}
The quantity of BHAD, despite its importance, is debatable as X-ray measurements are often significantly lower than theoretical predictions, particularly at high redshift ($z\gtrsim 3$) where detections are less complete. 
In this work, we constrain BHAD at $z=1.5$--5 using a novel method, from a sample of $z=0.7$--1.5 bulge-dominated galaxies (\S\ref{sec:data}).  
Our BHAD estimation \fst{(dominated by cold accretion; \S\ref{sec:intro})} is based on the BHAR-SFR correlation among bulge-dominated galaxies \citep{yang19} and the galaxy SFHs derived from their broad-band photometry and grism spectroscopy from \hst/WFC3 observations in the CLEAR survey.  

Our estimated BHAD is consistent with both of the theoretical 
predictions and the X-ray measurements at $z\lesssim 3$ 
(Fig.~\ref{fig:bhad}). 
At $z\approx 4$--5, our BHAD agrees with simulations, but it is higher than the X-ray results (by $\approx$3--10 times at $z=4$--5). 
After considering several causes of this discrepancy (\S\ref{sec:err}), we argue that it stems from two possibilities.  Either (1) there exists a large population of heavily obscured  Compton-thick AGN at $z \gtrsim 4$ current missed in \xray\ surveys (see \S\ref{sec:argue}), or (2) the BHAR-SFR relation begins to break down at $z \gtrsim 2.5$, which could result from a significant drop in the frequency of bulge-dominated galaxies. Either scenario can be tested with future observations.

\fst{The high-$z$ Compton-thick AGNs likely have strong IR emission.
In the future, \jwst\ and \origins\ will be able to detect dozens of high-$z$ Compton-thick AGNs in their deep broad-band imaging surveys, if this heavily obscured population is mainly responsible for the discrepancy between our BHAD and the X-ray results (see \S\ref{sec:argue}).
Another way is to identify high-$z$ AGNs with narrow emission lines (but this requires spectroscopy). 
For example, the high-$z$ version of BPT diagram \citep{baldwin81}, which is often used to classify AGNs vs. star-forming galaxies in the local universe, will be available from \jwst\ observations.
There are also some AGN-sensitive lines such as \nev\ and \oiv\ observable by \origins\ at $z\approx 4$--5 \citep[e.g.,][]{satyapal20}.
}
It will be interesting to further study the Lyman continuum escape fraction ($f_{\rm esc}$) of the \jwst/\origins-detected Compton-thick AGNs. 
We expect that $f_{\rm esc}$ to be low (nearly zero) considering the strong obscuration in \xray. 
But if this is not the case, then the high-$z$ Compton-thick 
AGNs could be an important source of cosmic reionization \citep[e.g.,][]{fan06, robertson15, finkelstein19}.

\jwst\ will also be able to test the frequency of bulge-dominated galaxies at high redshifts. 
Our BHAD estimation assumes that the high-$z$ progenitors of our 
sample remain bulge-dominated (\S\ref{sec:err}). 
This assumption could be impacted by  morphological transformation,
although observations find such transformation is unlikely prevalent at 
$z\lesssim 2.5$.
The currently available \hst\ $H$-band imaging is shifted into
rest-frame UV wavelengths at $z\gtrsim 2.5$, preventing reliable
morphological classifications \citep[e.g.,][]{conselice14, huertas_company15}. 
\jwst\ will overcome this issue by providing high-resolution 
imaging of wavelengths up to $\approx 5\ \mu$m (NIRCam).
If morphological transformation is responsible for our 
reported BHAD difference, \jwst\ will find that bulge-dominated 
galaxies are very rare, less than a few percent among massive 
galaxies at $z\approx 5$ (Fig.~\ref{fig:bhad_moduled}).  

In reality, we expect that the Universe will surprise us.  For example, it is reasonable to hypothesize that multiple effects  may be at play (including others not considered here).  We may  discover both a higher abundance of obscured AGN, evolution in the bulge-fraction of galaxies, and/or something entirely unexpected.  Regardless, the predictions are important as they provide a baseline, and then future studies will lead to an improved understanding of the history of BH accretion and the joint evolution of BH accretion and star-formation in galaxies.

\section*{Acknowledgments}
We thank the referee for helpful feedback that improved this work.
We thank our collaborators on the CLEAR project for valuable discussions and their work to provide a high-quality dataset.
In particularly, we thank Ivelina Momcheva, Raymond Simons, Gabriel Brammer, Yoshihiro Ueda, Tonima Tasnim Ananna, and James Aird for helpful discussions, suggestions, and/or providing relevant data.  
VEC acknowledges support from the NASA Headquarters under the Future Investigators in NASA Earth and Space Science and Technology (FINESST) award 19-ASTRO19-0122.
This work is based on data obtained from the Hubble Space Telescope through program number GO-14227. 
Support for Program number GO-14227 was provided by NASA through a grant from the Space Telescope Science Institute, which is operated by the Association of Universities for Research in Astronomy, Incorporated, under NASA contract NAS5-26555.
This work is supported in part by the National Science Foundation through grant AST 1614668. 
The authors acknowledge the Texas A\&M University Brazos HPC cluster and Texas A\&M High Performance Research Computing Resources (HPRC, http://hprc.tamu.edu) that contributed to the research reported here.

\software{
{\sc astropy} \citep[v4.2][]{astropy},
{\sc x-cigale} \citep{boquien19, yang20}.
}

\bibliography{all.bib}
\bibliographystyle{aasjournal}

\end{CJK*}
\end{document}